\documentclass[%
 reprint,
superscriptaddress,
showpacs,
 amsmath,amssymb,
 aps,
prl,
]{revtex4-1}

\usepackage{graphicx}
\usepackage{dcolumn}
\usepackage{bm}

\newcommand{\ket}[1]{\ensuremath{\left|#1\right\rangle}}

\begin{document}
\DeclareGraphicsExtensions{.pdf,.png,.jpg,.eps,.tiff}
\title{Ground-state cooling of a trapped ion using long-wavelength radiation}
\author{S. Weidt}
\affiliation{Department of Physics and Astronomy, University of Sussex, Brighton, BN1 9QH, UK}
\author{J. Randall}
\affiliation{Department of Physics and Astronomy, University of Sussex, Brighton, BN1 9QH, UK}
\affiliation{QOLS, Blackett Laboratory, Imperial College London, London, SW7 2BW, UK}
\author{S. C. Webster}
\affiliation{Department of Physics and Astronomy, University of Sussex, Brighton, BN1 9QH, UK}
\author{E. D. Standing}
\affiliation{Department of Physics and Astronomy, University of Sussex, Brighton, BN1 9QH, UK}
\author{A. Rodriguez}
\affiliation{Department of Physics and Astronomy, University of Sussex, Brighton, BN1 9QH, UK}
\author{A. E. Webb}
\affiliation{Department of Physics and Astronomy, University of Sussex, Brighton, BN1 9QH, UK}
\author{B. Lekitsch}
\affiliation{Department of Physics and Astronomy, University of Sussex, Brighton, BN1 9QH, UK}
\author{W. K. Hensinger}
\email{W.K.Hensinger@sussex.ac.uk}
\affiliation{Department of Physics and Astronomy, University of Sussex, Brighton, BN1 9QH, UK}
\begin{abstract}
We demonstrate ground-state cooling of a trapped ion using radio-frequency (RF) radiation. This is a powerful tool for the implementation of quantum operations, where RF or microwave radiation instead of lasers is used for motional quantum state engineering. 
We measure a mean phonon number of $\overline{n} = 0.13(4)$ after sideband cooling, corresponding to a ground-state occupation probability of 88(7)\%. After preparing in the vibrational ground state, we demonstrate motional state engineering by driving Rabi oscillations between the \ket{n=0} and \ket{n=1} Fock states. We also use the ability to ground-state cool to accurately measure the motional heating rate and report a reduction by almost two orders of magnitude compared to our previously measured result, which we attribute to carefully eliminating sources of electrical noise in the system. 

\end{abstract} 


\maketitle

Trapped atomic ions are a well established platform for the implementation of quantum computation \cite{Nigg, Home}, quantum simulation \cite{Kim, Lanyon} and frequency standards experiments \cite{Gill}. Realising such experiments often requires cooling the ions to below the Doppler limit. A very successful technique to achieve this is resolved sideband laser cooling, which was first implemented and used to reach the ground-state of motion in seminal experiments on a quadrupole transition using a very narrow-linewidth laser \cite{Diedrich} and on a Raman transition \cite{Monroe}. Further refinement of the technique has lead to a motional ground-state occupation probability of 99.9\% \cite{Roos}.

While the initial seminal examples of quantum state engineering and sub-Doppler cooling were performed using laser radiation, in recent years exciting new schemes for the creation of the required coupling between the internal state of an ion and its harmonic motion have been proposed in the pioneering work by Mintert and Wunderlich \cite{Mintert1} and Ospelkaus et al. \cite{Ospelkaus1} in which long-wavelength radio-frequency (RF) or microwave radiation is used instead. The successful demonstration of quantum operations such as high-fidelity single and multi-qubit gates as well as cooling to below the Doppler laser cooling limit to reach the ground-state of motion using RF or microwave radiation is expected to provide a powerful platform for scaling up a broad range of quantum technologies. In this approach lasers are only required for Doppler cooling, state detection and periodic dissipative reinitialisation of internal ionic states. Following the initial proposals, the first two-qubit gate using microwave radiation was implemented using near-field microwave gradients \cite{Ospelkaus} which was followed by the implementation of a two-qubit gate between nearest as well as non-nearest neighbours using microwaves in conjunction with a static magnetic field gradient \cite{Khromova1}. Microwave radiation has also been used to perform single-qubit gates with a fidelity far exceeding the minimum threshold for fault-tolerant quantum computing \cite{Brown, Harty}. 

Cooling an ion to the quantum ground-state of motion using RF or microwave radiation to drive motional sideband transitions, with lasers used purely for reducing entropy, would be highly desirable for a range of quantum state engineering experiments \cite{Harlander, Brown1, Barrett2, Brickman, Leibfried3, Haffner3, Riebe}. The Cirac-Zoller two-qubit gate fidelity reduces with increasing temperature \cite{03:schmidtb}, while for M{\o}lmer-S{\o}rensen and geometric phase gates cooling to the ground state reduces the effect timing and detuning imperfections have on the gate fidelity \cite{Hayes2}. Ground-state cooling is therefore an important resource for the realisation of fault-tolerant quantum computing with RF or microwave radiation. Initialising the motion to a pure quantum state also allows for non-classical motional states, such as Fock, coherent and squeezed states, to be created \cite{96:meekhof} and opens a pathway to a variety of experiments including precision measurements \cite{Chou} and quantum simulations \cite{12:schneider}.

So far, near-field microwave gradients have been used to sideband cool a cold high frequency radial mode to a mean phonon number of $\overline{n} = 0.6(1)$ \cite{Ospelkaus} and microwaves in conjunction with a static magnetic field gradient were used to sideband cool a lower frequency axial mode to $\overline{n} = 23(7)$ \cite{Khromova1}. 

Using such methods to further reduce the temperature and increase the ground state occupation probability remains a significant challenge.

In this manuscript we demonstrate ground-state cooling of a $^{171}$Yb$^{+}$ ion using a strong RF field in conjunction with a large static magnetic field gradient. 
Laser radiation is used only for initial Doppler cooling, repumping during
sideband cooling and preparation and detection of the ion's internal
state. We use two microwave fields to dress three of the atomic levels, creating a dressed-state qubit that has a long coherence time, yet is sensitive to magnetic field gradients, allowing spin-motion couplings to be realised. The use of the dressed-state qubit enhances the efficiency of population transfer for each sideband pulse in the cooling sequence, allowing lower final temperatures to be reached.
Following sideband cooling we measure a final phonon number of $\overline{n} = 0.13(4)$ which corresponds to a ground-state population probability of 88(7)\%. We combine the ability to prepare the vibrational Fock state $\ket{n=0}$ with the significant increase in coherence time of our dressed system to observe sideband Rabi oscillations which last for more than 10 ms and use our ground-state cooling method to measure the motional heating rate of our ion trap. 

A segmented macroscopic linear Paul trap with an ion-electrode distance of 310 $\mu$m is used to confine a single $^{171}$Yb$^{+}$ ion \cite{Lake}. Once the ion is trapped it is Doppler laser cooled on the $^{2}$S$_{1/2}\leftrightarrow$ $^{2}$P$_{1/2}$ transition using near-resonant 369 nm light. To form a closed cooling cycle any population in the $^{2}$D$_{3/2}$ state is transferred to the $^3[3/2]_{1/2}$ state using resonant light at 935 nm from which the population decays back to the $^{2}$S$_{1/2}$ state. By tuning the 369 nm light to be resonant with the $^{2}$S$_{1/2}$, F=1 $\leftrightarrow$ $^{2}$P$_{1/2}$, F=1 transition, the $\ket{0}\equiv$ $^{2}\text{S}_{1/2}\ket{F=0}$ state is prepared with near unit fidelity in $<$ 10 $\mu$s. State detection is achieved using a fluorescence threshold technique whereby light is tuned to be resonant with the $^{2}$S$_{1/2}$, F=1 $\leftrightarrow$ $^{2}$P$_{1/2}$, F=0 cycling transition \cite{Webster}. 

In order to obtain a coupling between the motion of the ion and RF radiation, we apply a large static magnetic field gradient to the ion using four permanent rare earth SmCo magnets integrated close to the ion trap. The static magnetic field gradient has been measured to be 23.6(3) T/m \cite{Lake}. By driving transitions within the ground-state with $\Delta m_\text{F} = \pm 1$ using RF radiation a coupling strength is obtained which scales with an effective Lamb-Dicke parameter (LDP) $\eta_{\rm eff} = z_0\mu_B\partial_zB/\hbar\nu_z$ \cite{Mintert1}. Here $\partial_zB$ is the static magnetic field gradient, $\nu_z$ is the axial secular frequency, $\mu_B$ is the Bohr magneton and $z_0=\sqrt{\hbar/2m\nu_z}$ is the spatial extent of the ground-state wave function. For a measured axial secular frequency of $\nu_z/2\pi = 426.7(1)$ kHz, which is used in this work, we obtain $\eta_{\rm eff} = 0.0064$. The ion is slightly displaced from the magnetic field null to give a magnetic field offset at the ion of 10.5 G which lifts the degeneracy of the three $^{2}$S$_{1/2}$, F=1 Zeeman-levels $\ket{0'}\equiv$ $^{2}$S$_{1/2}\ket{F=1, m_F=0}$, $\ket{+1}\equiv$ $^{2}$S$_{1/2}\ket{F=1, m_F=+1}$ and $\ket{-1}\equiv$ $^{2}$S$_{1/2}\ket{F=1, m_F=-1}$ by $\approx$ 14.6 MHz and results in a second-order Zeeman shift which separates the $\ket{0'}\leftrightarrow \ket{+1}$ transition and the $\ket{0'}\leftrightarrow \ket{-1}$ transition by 34 kHz \cite{Webster}.

\begin{figure}[tb]
\centering
\includegraphics[width=0.48\textwidth]{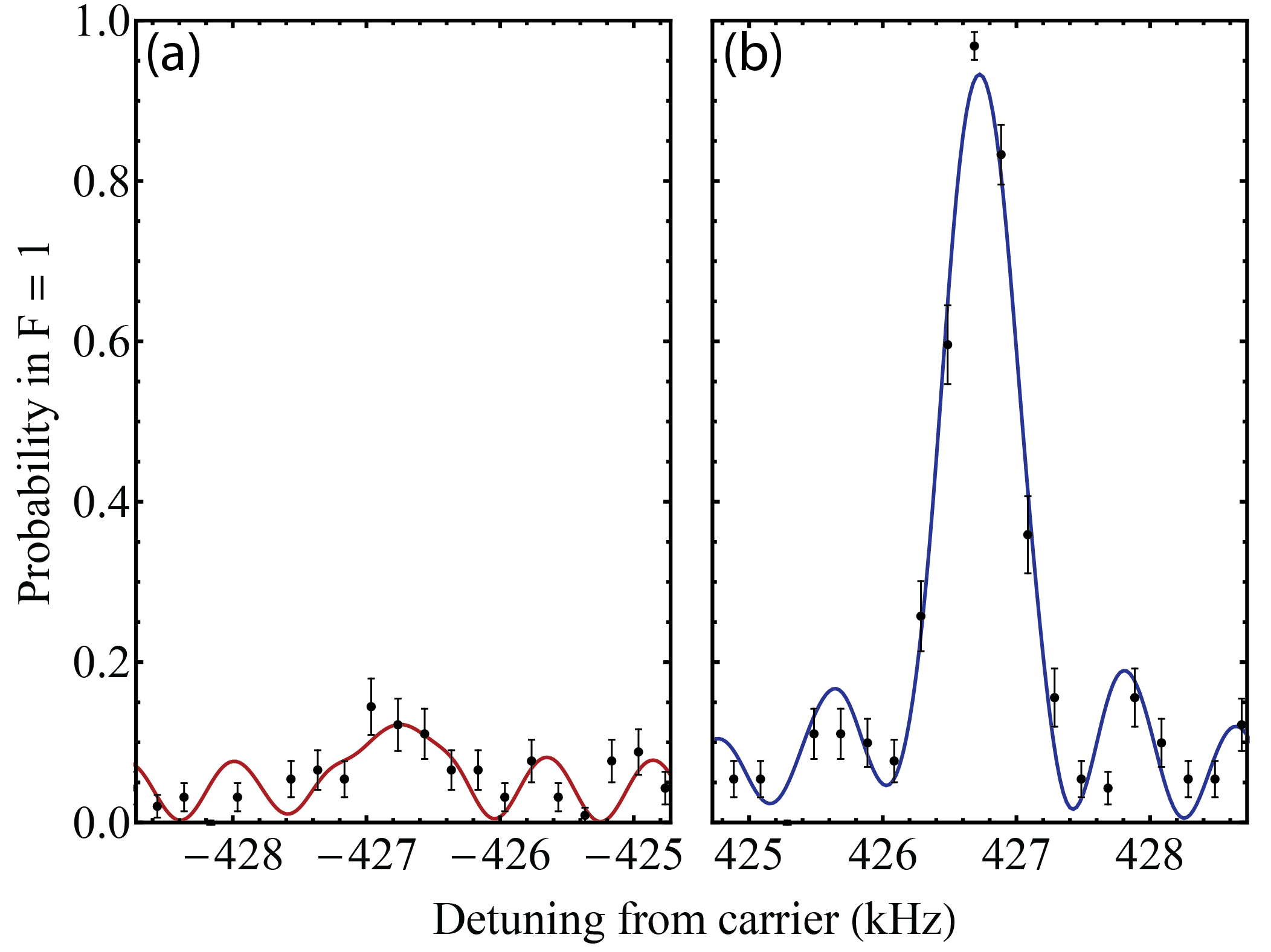}
\caption{Population in F = 1 after (a) a frequency scan over the red sideband and (b) a frequency scan over the blue sideband, both after sideband cooling. The red and blue lines are the result of a numerical simulation of the system where we have set $\nu_z/2\pi = 426.7$ kHz, $\Omega_\text{dr}/2\pi = 32$ kHz, $\Omega/2\pi = 61.2$ kHz and a probe pulse time of 1270 $\mu$s with $\bar{n}$ as the only free parameter. A least-squares fit gives $\bar{n} = 0.13(4)$, corresponding to a ground state occupation probability of $p_0 = 0.88(7)$.}

\label{sideband_cooling}
\end{figure}

The requirement to use states with different magnetic moments for spin-motion coupling with a static magnetic field gradient limits the coherence time, as the system is sensitive to ambient magnetic field fluctuations \cite{Khromova1, Lake}. The coherence time of our bare magnetic field sensitive system is $\approx$ 1 ms which is on the same order of magnitude as the ground-state sideband $\pi$-time used in our experiment and therefore limits the efficiency of the cooling process. To significantly increase the coherence time of our system, we apply a pair of dressing fields using microwaves near 12.6 GHz to the ion which resonantly couple the two magnetic field sensitive F=1 states $\ket{+1}$ and $\ket{-1}$ with the first-order magnetic field insensitive F=0 state $\ket{0}$ with equal Rabi frequency $\Omega_\text{dr}$ \cite{Timoney, Webster, Randall2}. This gives three dressed-states, one of which $\ket{D}=(\ket{+1}-\ket{-1})/\sqrt{2}$ can be combined with $\ket{0'}$ to form an effective two-level system that is resilient to noise in the magnetic field \cite{Timoney, Webster, Randall2}. Preparation and detection of the dressed-state system is achieved using the method described in Ref. \cite{Randall2}. For preparation, a microwave $\pi$-pulse transfers the population to $\ket{0'}$ after the ion has been optically pumped into the $\ket{0}$ state. The microwave dressing fields are then applied instantaneously, after which an RF field resonant with the $\ket{0'}\leftrightarrow \ket{+1}$ transition is used to couple $\ket{0'}$ and $\ket{D}$ \cite{Webster, Randall2}. Detuning the RF field either side of the carrier transition by the motional trap frequency then allows the coupling to motional sideband transitions, analogous to an undressed two-level system, however also resilient to magnetic field fluctuations. To detect the final state, the dressing fields are turned off, after which a microwave $\pi$-pulse swaps population between $\ket{0'}$ and $\ket{0}$. The fluorescence threshold detection technique can now be used to distinguish between population in $^{2}$S$_{1/2}$, F=1 and $\ket{0}$, corresponding to the effective two-level system $\ket{D}$ (absent any decoherence during the dressing) and $\ket{0'}$ respectively.

To ground-state cool the ion, a pulsed sideband cooling technique \cite{Monroe} is used that consists of a repeated sequence which includes driving a motional sideband transition and a dissipative repumping process to reset the ion's internal state.
The ion, starting in the state \ket{0'}, is dressed with microwave fields as described above. 
An RF field resonant with the red sideband transition, detuned from the carrier transition $\ket{0'}\leftrightarrow$ $\ket{D}$ by the secular frequency $\nu_z$, is then applied for a time $t$. A transition of the ion's internal state to \ket{D} reduces the ion's motional energy by $\hbar\nu_z$. To reset the ion's internal state to $\ket{0'}$, the dressing fields are turned off, a microwave $\pi$ pulse applied to swap the populations between $\ket{0'}$ and $\ket{0}$, and the ion optically pumped into \ket{0} using 369 nm laser light. Finally, a second $\pi$ pulse transfers the ion to \ket{0'}, ready to start the sequence again. To optically pump an ion from one of the F=1 states to \ket{0} requires on average $\sim3$ photons to be scattered, causing heating. The first $\pi$ pulse, before the optical pumping step, ensures that this photon scattering occurs only if the ion was transferred to \ket{D} by the RF pulse. By repeating this sideband cooling sequence multiple times, the ion's temperature is reduced towards $\bar{n}=0$. 

\begin{figure}[tb]
\centering
\includegraphics[width=0.45\textwidth]{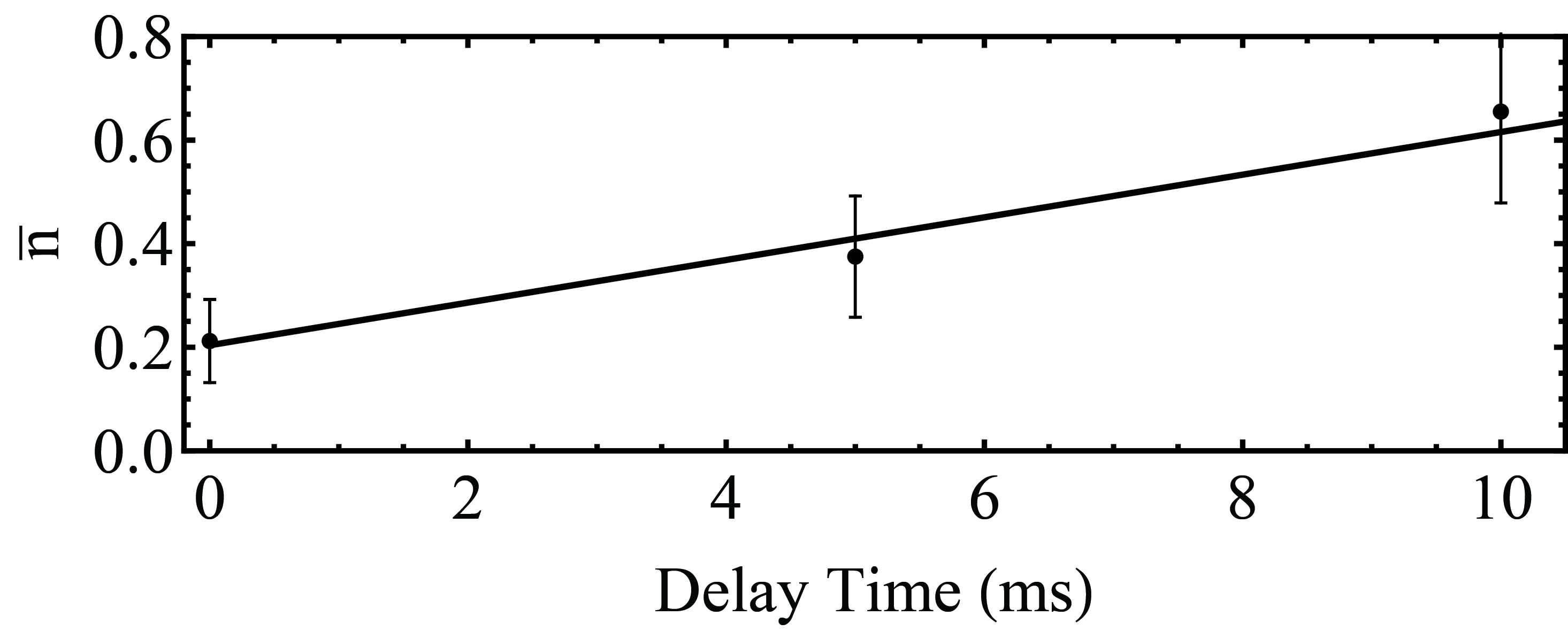}
\caption{To measure the heating rate $\dot{\bar{n}}$ a variable delay is introduced after sideband cooling. For each of the three delay times a scan over the red and blue sidebands is taken, as in Fig. \ref{sideband_cooling}, allowing $\bar{n}$ to be determined. The black line is a fit to the data which gives a heating rate of $\dot{\bar{n}} = 41(7)$ s$^{-1}$, or 1 phonon in 24 ms, for the axial secular frequency of $\nu_z/2\pi = 426.7(1)$ kHz.}
\label{heating_rate}
\end{figure}

Before sideband cooling, we determine the initial temperature of the ion after Doppler cooling by measure the population in \ket{D} as a function of the time that we resonantly drive the red sideband transition for. 
The probability to make a transition from $\ket{0'}$ to $\ket{D}$ for a resonant sideband transition is given by
\begin{equation}\label{eq:Psb}
P_{\pm}(t) = \sum_{n=0}^\infty p_n \frac{1}{2}\left(1-\cos(\Omega_{n,n\pm 1}t)\right),
\end{equation}
where the $-$($+$) sign is used for a red (blue) sideband transition respectively. Here $p_n$ is the population in the $n$th motional state and, to first-order in $\eta_{\rm eff}$, $\Omega_{n,n-1} = \eta_{\rm eff}\Omega\sqrt{n}$ is the Rabi frequency for the $\ket{0',n} \leftrightarrow \ket{D,n-1}$ transition, while $\Omega_{n,n+1} = \eta_{\rm eff}\Omega\sqrt{n+1}$ is the Rabi frequency for the $\ket{0',n} \leftrightarrow \ket{D,n+1}$ transition. $\Omega$ is the carrier Rabi frequency and for an ion in a thermal distribution of motional states characterised by a mean phonon number $\bar{n}$, the motional state populations are given by $p_n = (1/(\bar{n}+1))(\bar{n}/(\bar{n}+1))^n$.
By fitting the resultant data to Eq. \ref{eq:Psb} we find the initial temperature after 4 ms of Doppler cooling to be $\bar{n} = 65(5)$. This is approximately 3 times the Doppler cooling limit of $\bar{n} = 23$.

Following Doppler cooling to $\bar{n} = 65(5)$, the sideband cooling sequence method explained above is experimentally implemented as follows. The Rabi frequencies of the two dressing fields have been independently measured to be $\Omega_\text{dr}/2\pi = 32$ kHz.  The RF field has a carrier Rabi frequency of $\Omega/2\pi = 61.2$ kHz and is made resonant with the red sideband of motion by detuning from the carrier by the axial secular frequency of $\nu_z/2\pi = 426.7(1)$ kHz. For the repumping step, the resonant microwave $\pi$-pulses to swap population between $\ket{0}$ and $\ket{0'}$ each last 14 $\mu$s, and the 369 nm light for optical pumping into $\ket{0}$ is applied for 6 $\mu$s. We apply a total of 500 repetitions of the sideband cooling sequence, where for each repetition the RF sideband pulse is applied for an increasing length of time. The sideband pulse times are set to be $t_n = \pi/\Omega_{n,n-1}$ for each $n$ level in turn, starting from $n = 500$, giving a total sideband cooling time of 71 ms. This could be reduced by more than an order of magnitude by reducing the initial temperature, and increasing the Rabi frequencies of our driving fields and the magnetic field gradient. 

To measure the final temperature after sideband cooling, an RF probe pulse is applied after preparing in $\ket{0'}$ and applying the dressing fields. The frequency of the RF field is scanned over both the red and blue sideband. The ratio of probabilities to make a transition from $\ket{0'}$ to $\ket{D}$ when resonant with the red or blue sideband can be shown from Eq. \ref{eq:Psb} to be $r = P_{-}(t)/P_{+}(t) = \bar{n}/(\bar{n}+1)$, assuming the ion is in a thermal state \cite{Monroe}. 
After the sideband cooling sequence the frequency of the RF probe field is scanned over the red and blue sidebands. The result of this scan, for an RF pulse time of 1270 $\mu$s, is shown in Fig. \ref{sideband_cooling} (a) and (b) for the red and blue sideband respectively. From this we extract a final mean phonon number of $\bar{n} = 0.13(4)$, corresponding to a ground-state occupation probability of $p_0 = 0.88(7)$. As explained below, this result is consistent with the minimum temperature estimated when taking into account effects from heating.

\begin{figure}[tb]
\centering
\includegraphics[width=0.45\textwidth]{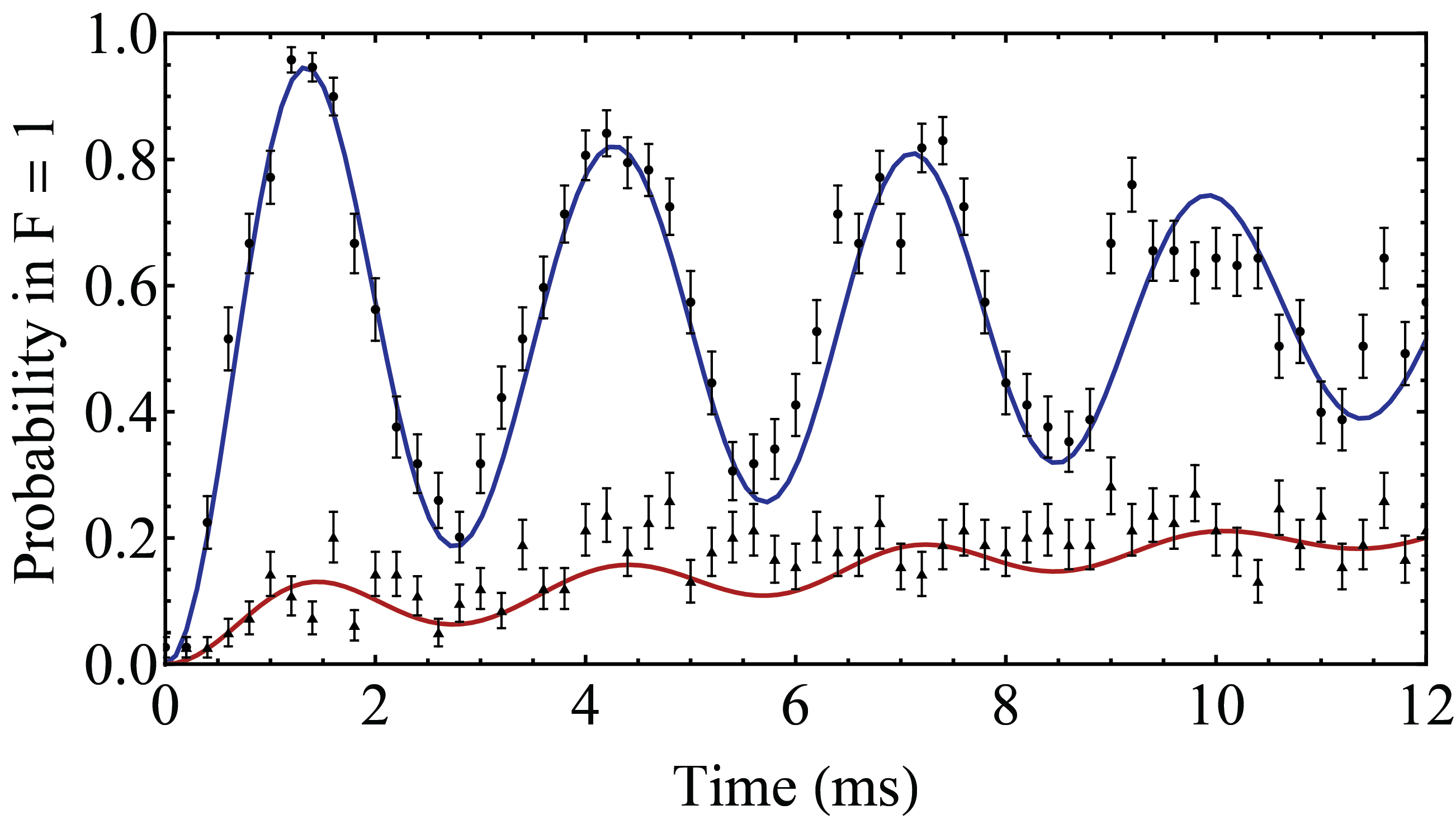}
\caption{Population in F = 1 after ground state cooling and applying a RF pulse for a variable time, resonant with the blue (circles) and red (triangles) motional sidebands of the $\ket{0'} \leftrightarrow \ket{D}$ transition. Coherent Rabi oscillations between $\ket{0',n=0}$ and $\ket{D, n=1}$ can be observed lasting for over $10$ ms, which is an order of magnitude longer than gate times achievable in this setup. The blue and red solid lines are the results of integrating a master equation describing the dynamics of a two-level atom coupled by either the red or blue sideband to a harmonic oscillator mode undergoing heating. The heating rate and initial thermal state are set to the measured values of $\dot{\bar{n}} = 41$ s$^{-1}$ and $\bar{n}=0.13$ respectively, and the ground state sideband Rabi frequency is set to $\eta_{\rm eff}\Omega/2\pi = 0.35\,{\rm kHz}$.}
\label{sideband_rabi_flop}
\end{figure}

A limiting factor of the maximum achievable ground-state population probability is the heating rate of the motional mode being cooled. The heating rate of an ion in our trap has previously been measured and a scaled spectral noise density of $\nu_z S_E(\nu_z) = 4\dot{\bar{n}}\hbar m \nu_z^2/e^2 = 2.3(6) \times 10^{-4}$ V$^2$m$^{-2}$ has been calculated \cite{McLoughlin2}. This would correspond to a heating rate of 6700 s$^{-1}$ at a secular frequency of $\nu_z/2\pi = 426.7(1)$ kHz. 
Methods to significantly reduce the heating rate include cooling of the trap electrodes \cite{Deslauriers1} and the use of a sputter gun \cite{Hite} or a high intensity pulsed-laser \cite{Allcock2} to clean the surface of the trap. These methods are however only useful if no other dominating sources of heating are present. Electrical noise is such a potential dominating source of heating \cite{Brownnutt, Private} and we have acted to reduce it in several ways. We  developed a new low noise multi-channel voltage supply for the static voltage trap electrodes. We also removed other sources of electromagnetic noise in the vicinity of the experiment by replacing noise-inducing electronics and use a well separated ground for relevant low-noise electronics.

To measure the new heating rate of the ion we modify the sideband cooling sequence experiment described above to include a variable delay time. The delay is inserted after the sideband cooling but before the RF probe pulse, during which the ion heats at a rate $\dot{\bar{n}}$. By performing the experiment for delay times of 0 ms, 5 ms and 10 ms, a heating rate can be extracted. The measured temperature for each delay time is shown in Fig. \ref{heating_rate}. A fit to the data gives a heating rate of 41(7) s$^{-1}$, which corresponds to a gain of 1 phonon every 24 ms. This heating rate gives a scaled electric-field noise density of $\nu_z S_E(\nu_z) = 1.4(2) \times 10^{-6}$ V$^2$m$^{-2}$, which is a reduction of more than two orders of magnitude compared to our previously measured value and reiterates the importance of carefully controlling electrical noise present in the laboratory.

The ability to ground-state cool in conjunction with the long coherence time offered by our dressed-system allows us to demonstrate simple motional state engineering.  We sideband cool the axial mode of motion using the method described above and prepare the Fock state $\ket{n=0}$. We then apply a RF field resonant with the blue sideband of motion ($\ket{0',n=0}\leftrightarrow \ket{D,n=1}$) for an increasing time, resulting in Rabi oscillations with frequency $\eta_\text{eff}\Omega/2\pi = 0.35$ kHz, coherently manipulating the phonon state. This is shown in Fig. \ref{sideband_rabi_flop} and demonstrates that Rabi oscillations are maintained for over 10 ms, an order of magnitude longer than achievable two-qubit gate times currently envisioned to be implemented in this system. We also apply a RF field resonant with the red sideband for an increasing time in a separate experiment, which is also shown in Fig. \ref{sideband_rabi_flop}. A theoretical line for both sets of data is overlaid, which is the result of numerically integrating a master equation incorporating the effect of heating. It can be seen that using the measured heating rate, the data closely matches the simulated curves.

We have experimentally demonstrated ground-state cooling of an ion by coupling an effective two-level microwave-dressed system to its motion using RF radiation in conjunction with a large static magnetic field gradient. We also show coherent manipulation of the motional state, by demonstrating multiple Rabi oscillations on the blue sideband which corresponds to the repeated exchanges of excitation between the internal and motional states of the ion. Furthermore, we have measured a reduction of the motional heating rate of the ion by almost two-orders of magnitude, which we attribute to careful minimisation of external electric field noise. By demonstrating ground-state cooling using rf radiation to drive the sideband transitions, we complete a toolbox of techniques for quantum state engineering using RF or microwave radiation.

Laser radiation is used only for detection, Doppler cooling and repumping, removing the need for highly stable laser sources or laser beams in a Raman configuration. Instead of addressing individual ions using multiple spatially separated tightly focused laser beams, ions in a magnetic field gradient are spectrally differentiable and can therefore be addressed with global control fields \cite{Piltz2}. Local microwave fields applied via on-chip waveguides could also be used for this task \cite{Warring, Craik}. Our method can be scaled by allowing the application of an offset magnetic field to individual entanglement zones, thereby allowing ions in arbitrary zones to be tuned in and out of resonance with globally applied sets of microwave and RF fields where each set corresponds to a particular type of gate to be performed \cite{Lekitsch}. 

\section{Acknowledgements}

This work is supported by the U.K. Engineering and Physical Sciences Research Council  [EP/G007276/1, EP/E011136/1, the UK Quantum Technology hub for Networked Quantum Information Technologies (EP/M013243/1), the UK Quantum Technology hub for Sensors and Metrology (EP/M013243/1)], the European Commission’s Seventh Framework Programme (FP7/2007-2013) under Grant Agreement No. 270843 (iQIT), the Army Research Laboratory under Cooperative Agreement No. W911NF-12-2-0072, the US Army Research Office Contract No. W911NF-14-2-0106 and the University of Sussex. The views and conclusions contained in this document are those of the authors and should not be interpreted as representing the official policies, either expressed or implied, of the Army Research Laboratory or the U.S. Government. The U.S. Government is authorized to reproduce and distribute reprints for Government purposes notwithstanding any copyright notation herein.

\bibliography{ReportBib}

\end{document}